\documentclass[aps,prl,twocolumn,superscriptaddress]{revtex4-1}
\usepackage{amsmath}% Include figure files
\usepackage{graphicx}% Include figure files
\usepackage{dcolumn}% Align table columns on decimal point
\usepackage{bm}% bold math

\begin{document}

%\preprint{APS/123-QED}

\title{Positronium in MOFs: the Atom out of the box}% Force line breaks with \\

\author{P.~Crivelli}
\affiliation{Institute for Particle Physics, ETH Zurich, Switzerland}
\email[]{crivelli@phys.ethz.ch}
\author{D.~Cooke}
\affiliation{Institute for Particle Physics, ETH Zurich, Switzerland}
%\email[]{cooke@phys.ethz.ch and crivelli@phys.ethz.ch}
%
\author{B. Barbiellini}
\affiliation{Department of Physics, Northeastern University, Boston, Massachusetts 02115, USA}

\author{B. L. Brown}
\affiliation{Department of Physics, Marquette University, Milwaukee, USA}
\author{J.~I.~Feldblyum}
\affiliation{Department of Chemistry, University of Michigan, 48109, USA}
\affiliation{Macromolecular Science and Engineering, University of Michigan, 48109-2136, USA}
%\affiliation{Department of Chemistry, University of Michigan, 930 North University Street, Ann Arbor, Michigan 48109, USA}
%\affiliation{Macromolecular Science and Engineering, University of Michigan, 2300 Hayward Avenue, Ann Arbor, Michigan 48109-2136, USA}
%
\author{P.~Guo}
%\affiliation{Department of Chemistry, University of Michigan, 930 North University Street, Ann Arbor, Michigan 48109, USA}
\affiliation{Department of Chemistry, University of Michigan, 48109, USA}

\author{D.~W.~Gidley}
\affiliation{Department of Physics, University of Michigan, 48109, USA}
%\affiliation{Department of Physics, University of Michigan, 450 Church Street, Ann Arbor, Michigan 48109, USA}
%
\author{L.~Gerchow}
\affiliation{Institute for Particle Physics, ETH Zurich, Switzerland}
\author{A.~J.~Matzger}
\affiliation{Department of Chemistry, University of Michigan, 48109, USA}
%\affiliation{Department of Chemistry, University of Michigan, 930 North University Street, Ann Arbor, Michigan 48109, USA}
%%
%
\date{\today}% It is always \today, today,

%\clearpage
\begin{abstract}
Recently, evidence for positronium (Ps) in a Bloch state in self-assembled metal--organic frameworks (MOFs) has been reported [Dutta et al., Phys. Rev. Lett. 110, 197403 (2013)]. In this paper, we study Ps emission into vacuum from four different MOF crystals: MOF-5, IRMOF-8, ZnO$_4$(FMA)$_3$ and IRMOF-20. Our measurements of Ps yield and emission energy into vacuum provide definitive evidence of Ps delocalization. We determine with a different technique Ps diffusion lengths in agreement with the recently published results. Furthermore, we measure that a fraction of the Ps is emitted into vacuum  with a distinctly smaller energy than what one would expect for Ps localized in the MOFs' cells. We show that a calculation assuming Ps delocalized in a Kronig--Penney potential reproduces the measured Ps emission energy.
% We show how the population in the Bloch state can be controlled by tuning the initial positron beam energy. 
%We will show that this is a first step towards a Ps condensate.
%A more sophisticated model could help to shed some light on the debated question of the aperture size in MOFs. This parameter is relevant to model diffusion in these frameworks and thus its knowledge could be useful for their applications as mebranes or in catalysis processes.
\end{abstract}

%\end{frontmatter}
\maketitle

% Introduction **********************

Positronium (Ps), the bound state of the electron and its anti-particle (the positron), is the lightest atom in nature, which has inspired many fascinating studies \cite{Deutsch1}-\cite{gidley1} in fundamental and applied research. In particular, Ps delocalization in a Bloch state is a rare phenomenon that can be observed only in few crystals such as quartz. Historically, Brandt et al. \cite{brandt}  discovered a spectacular fine structure in the electron-positron momentum density of quartz while Greenberger et al. \cite{Greenberger} demonstrated that this feature is the manifestation of Ps formation in the form of a Bloch state.  The same phenomena was also observed in alkali halides \cite{brandt,inoue}.
In both quartz and halides, the positron wave function strongly overlaps with electrons outside the Ps. Therefore, the annihilation with electrons having an anti-parallel spin reduces the ortho/triplet Ps lifetime to about only few hundreds of ps \cite{nagai}. However, to comprehensively study and use Ps Bloch waves it is highly desirable to expand this lifetime.\par
Evidence for Ps Bloch states living tens of ns in self-assembled metal--organic frameworks (MOFs) has recently been reported \cite{Dutta2013}. MOFs are formed from the self-assembly of metal atoms or clusters linked by organic ligands (``linkers'') in highly regular structures with nanometer lattice size. The results are materials with extremely porous lattices and having surface areas up to and exceeding 5000 m$^2$/g \cite{janiak,farha}. Development has been driven by interest in industrial application of such highly sorbent materials to catalysis and gas (particularly hydrogen) storage \cite[e.g.][]{Makal2012}. By studying the emission probability of Ps escaping from large MOF grains Dutta et al. \cite{Dutta2013} deduced record-long Ps diffusion lengths that increased at lower temperatures, consistent with Ps existing primarily as a delocalized (Bloch) state in the lattice.\par
\begin{figure}[!h]
\centering
\includegraphics[width=0.5\columnwidth]{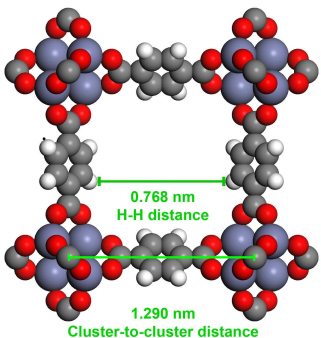}
\vspace{0.cm}
%\caption{Schematic representation of Ps (top) and cluster-to-cluster and H-H distances in MOF-5.}
\caption{Cluster-to-cluster and H-H distances in MOF-5.}
\label{fig:MOF5}
\end{figure}

In this letter, we definitely demonstrate the existence of a long lived Ps Bloch wave in MOF by studying the energy spectrum of Ps emitted into vacuum from four MOFs with different lattice parameter: ZnO$_4$(FMA)$_3$ (hereafter, FMA) \cite{fma}, MOF-5  \cite{mof5}, IRMOF-8 \cite{irmof8}, and IRMOF-20 \cite{irmof20}. The emission spectrum provides a direct view of the state energies available to Ps in the lattice. The Bloch state has a distinguishable energy that depends on each MOF's lattice size. All these MOFs are based on the ZnO$_4$ cluster, and thus form an \textit{isoreticular} series (hence IRMOF), that is, they possess the same net though the dimensions change according to the length of the chosen linker. The cluster-to-cluster and closest hydrogen--hydrogen distances (see Fig. \ref{fig:MOF5}) are reported in Table \ref{tbl1}.  The H--H distance is the distance from the closest H atoms on opposite sides of the framework having subtracted the van der Waals radii ($2\times1.2$ \AA).

\begin{table}
\caption{Cluster-to-cluster and H--H distances for the 4 samples studied in this paper.}\label{tbl1}
\begin{ruledtabular}
\begin{tabular}{cccc}
Sample & Density (g/cm$^3$)&Cluster--cluster (nm) & H--H (nm)\\\hline
IRMOF-20 & 0.511& 1.469 & 0.944\\
IRMOF-8 & 0.448 & 1.505 & 0.861\\
MOF-5   & 0.593& 1.290 & 0.768\\
FMA & 0.812 & 1.082 & 0.635\\ 
\end{tabular} 
\end{ruledtabular}
\end{table}

% 
%We will show that our measurement not only provide additional evidence for Ps in a Bloch state but also a way to control the occupation of the population in the Bloch state.
% and that this is a first step towards a Ps condensate. 

% Method ********************************

For this study we used the ETHZ slow e$^+$ beam \cite{Alberola2006,oPsTOF}. The slow positrons from the beam are accelerated to 1--20 keV towards the MOF target. After implantation, the e$^+$ can either form Ps, (o-Ps or p-Ps), or annihilate into $2\gamma$-rays directly. Positron implantation is tagged by detection of secondary electrons (SE), generated when the e$^+$ hit the target, with a microchannel plate (MCP). This provides the start time ($t_0$) to build the time distributions used to measure the yield and energy of Ps emitted into vacuum. The stop is given by the annihilation photons detected with three different spectrometers mounted around the target region: a large solid angle array of BGO crystals to reconstruct the fraction and the lifetime of Ps emitted in vacuum ($\sigma$=2.5 ns); a BaF$_2$ crystal to measure the lifetime and fraction of the Ps in the target($\sigma$=0.6 ns); and a time-of-flight (TOF) spectrometer made of one BGO crystal behind a lead collimator to determine the emission energy of Ps in vacuum ($\sigma$=2.5 ns). In order to maximize the signal-to-background ratio, the width of the collimator slit is set to 5 mm and its center was placed at a distance of 20 mm from the target. More details concerning the experimental setup can be found in \cite{oPsTOF}.\par
% Discussion (emission energy, diffusion etc.) *********************************
The MOF crystals, synthesized at the University of Michigan, typically have grain sizes of 300~$\mu$m, cells of 1-1.5 nm (see Table \ref{tbl1}) and densities of 0.45-0.81 g/cm$^3$. The mean e$^+$ implantation depth, estimated using a Mahkovian profile \cite{Makhovian}, is of the order of 0.5~$\mu$m for a $e^+$ implantation energy of 5~keV, rising to $\sim$3 $\mu$m for 15 keV. The positrons implanted in the MOF target rapidly thermalize through collision with molecules of the MOF framework; a fraction of these form Ps.
Given that for MOF-5, one has $\sim$75\% Ps formation (ortho + para) suggests that the Ps might indeed form directly by either Ore or spur processes in the almost 80\% porous volume as it would in a molecular gas that just happens to be arrayed in a regular lattice (Ps formation in low density gases can be nearly 100\% efficient).   
% Those atoms diffuse until they are ejected in the pores. Alternatively, the positron exiting the surface can capture an electron and form Ps there, the resulting atom subsequently possesses significant kinetic energy.\par

MOFs have framework sizes of the order of 1 nm, which is comparable with the Ps de Broglie wavelength for the energy range of interest, so Ps diffusion has to be treated quantum mechanically. 
%Due to the remarkable open cubic structure of MOFs, one would expect the calculations for a particle in a 3D box to be an good approximation of Ps behavior in these crystals. 
\begin{figure}[!h]
\centering
\includegraphics[width=0.6\columnwidth]{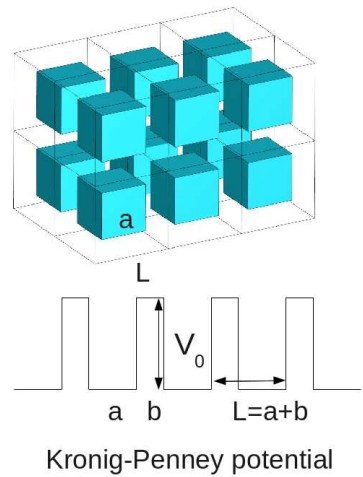}
\vspace{0.cm}
\caption{Muffin tin geometry of the MOFs and Kronig-Penney potential.}
\label{fig:Muffin}
\end{figure}
A single cell of MOF (side $L$) is connected to neighbouring cells via apertures of width $a$, where $a < L$. A particle entering an aperture must give up longitudinal energy in order to propagate through it. So in a narrow aperture, modes with transverse energy $V_0$ larger than the total energy of the particle do not propagate and decay exponentially fast as electromagnetic waves in a waveguide. This brings us to the result \cite{Batista} that a quantum mechanical transport of a particle through a narrow aperture is equivalent to the one dimensional propagation of a quantum particle across a potential barrier with a potential height $V_0$:
\begin{align}
V_0 = \frac{\hbar^2\pi^2}{ma^2}.\label{eq:V0}
\end{align}
Therefore, we can pass from the description of MOF large empty sites connected by channels with an aperture $a$ (see Fig. \ref{fig:MOF5}) to empty boxes with side $a$ whose separating walls contain the confinement potential $V_0$ as shown in Fig. \ref{fig:Muffin}. 
Depending on the linker configuration the side of the box might differ from the aperture, however we believe this is a reasonably good approximation for the MOFs we studied. 
In this picture, the `pore' for Ps is thus defined by a finite (square) well of width $a$ and height $V_0$. To calculate the bound states, one has to solve an equation for each direction $(x,y,z)$ given: 
\begin{eqnarray}\label{eq:MA}
 \sqrt{u_0^2-v^2} & = &\begin{cases} v \tan v & \mbox{(symmetric case)} \\ 
  -v \cot v & \mbox{(antisymmetric case) }\end{cases}
\end{eqnarray}
where  $v=a k/2$ and $u_0^2=m a^2 V_0/2\hbar^2=\pi^2/2$. In this case there are two solutions: $v_1=1.06$ and $v_2=2.01$. The energy levels of the 4 possible non-degenerate bound states in the 3D finite potential well are given by:  
\begin{equation}\label{eq:Elevels}
E_{ijk} = \frac{2 \hbar ^2}{m a^2}(v_i^2+ v_j^2+ v_k^2)~~{\mathrm{ for }}~~i,j,k \in\{1,2\}
\end{equation}
During diffusion, Ps can lose energy by phonon scattering, eventually becoming localized in single MOF cells and populating the discrete energy levels. At this point, one would expect that diffusion could only proceed further via tunnelling (as observed for muonium in porous silica thin films \cite{MuoniumPRL}). After sufficient thermalization time, Ps formed in the MOF should be emitted into vacuum with a discrete energy spectrum, corresponding to the energy levels of the MOF cell. The time Ps spends in the crystal is related to the e$^+$ implantation depth, so for e$^+$ implanted with greater energy, lower energy Ps is emitted. At room temperature, only the ground state should be populated, as in silica with 3--4 nm pores \cite{oPsTOF,Cassidy2010}, so for sufficient e$^+$ implantation energy, one should observe Ps only with energies in the range of 200-400 meV (we use the cluster-to-cluster distance of the MOFs reported in Table \ref{tbl1} to set this lower bound).\par

However, in addition to the different energy levels we measure that a fraction of Ps is emitted into vacuum with much lower energy. This can be understood if Ps falls into a delocalized (Bloch) state as a result of the periodic structure of MOFs. One can calculate the expected emission energy using a variety of `muffin tin' geometry for the potential where cubic boxes of size $a$ replace muffin tin spheres (see Fig. \ref{fig:Muffin}). The distance between boxes is the cluster-to-cluster distance (see Table \ref{tbl1}) $L$, therefore the width of the potential barrier (of height $V_0$) between the boxes is $b=(L-a)$. This model can be solved exactly, as in each direction $(x,y,z)$ it reduces to the well-studied Kronig-Penney problem:
\begin{align}
\frac{\gamma^2 - \beta^2}{2\beta \gamma}\sinh(\gamma b) \sin(\beta a) + \cosh(\gamma b) \cos(\beta a) = 
 \cos[k (a + b)]\label{eq:kp}
\end{align}
where $\gamma = \kappa \sqrt{V_0 - E}$, $\beta = \kappa \sqrt{E}$, $\kappa = \sqrt{(2 m)/\hbar^2}$ and $E$ the Ps energy.\par

The transition from localized to delocalized state could be explained by the Bondarev-Hyodo self-trapping mechanism \cite{bondarev}. The hot Ps by interacting with longitudinal acoustic phonons  produces small lattice distortions and generates a metastable self-trapped state for the Ps in the box. Eventually, when Ps cools down and the phonon population decreases it can jump from a localized metastable state to its most stable delocalized state. This can also be understood in terms of Anderson localization \cite{Anderson1958}. A hot Ps interacting with the lattice can easily emit phonons producing variations of on-site energy $\delta V$ associated with vibration of the box.  If $\delta V/BW >>1$ (where $BW$ is the bandwidth) Ps localizes \cite{baugher}. Only when Ps has dissipated all its energy to emit phonons such as $\delta V>BW$ its migration can be wavelike (in our case $BW\simeq 0.2$ eV for the lowest band). Therefore, higher lying Bloch states (i.e. excited bands) are probably too fragile to exist because of the decoherence by phonons: an intense level of scattering of an excited Ps with phonons would give the incoherent limit beyond which Bloch-wave propagation becomes inhibited. We thus do not include them in the analysis of our results.

% Results ********************************************

From the time distributions we determine the yield of Ps emitted into vacuum in the usual way performing a fit of the long exponential corresponding to the Ps lifetime in vacuum \cite{OurAPL}. We define the Ps vacuum yield as the probability of Ps emission into vacuum per implanted $e^+$. For more than 5 kV, Ps is assumed to be thermalized (see later) thus the diffusion length is constant. By fitting the vacuum yield curves we can determine the diffusion lengths of Ps in the various MOFs. These values are shown in Table \ref{tbl2}, and are in agreement with the ones extracted using a different experimental technique \cite{Dutta2013}. 

\begin{table}
\caption{Diffusion lengths and lifetime in the MOFs.}\label{tbl2}
\begin{ruledtabular}
\begin{tabular}{ccc}
Sample & Diffusion length ($\mu$m) & $\tau_{MOF}$ (ns)\\
\hline
IRMOF-20	& 3.0$\pm$0.2 	& 20$\pm$1\\
IRMOF-8		& 2.4$\pm$0.2 	& 18$\pm$1\\
MOF-5    	& 6.3$\pm$0.4	& 13$\pm$1 \\
FMA 		& 9.3$\pm$5.5 	& 10$\pm$1\\ 
\end{tabular} 
\end{ruledtabular}
\end{table}

\begin{figure}[h!]
%\begin{center}
\includegraphics[width=0.2\textwidth]{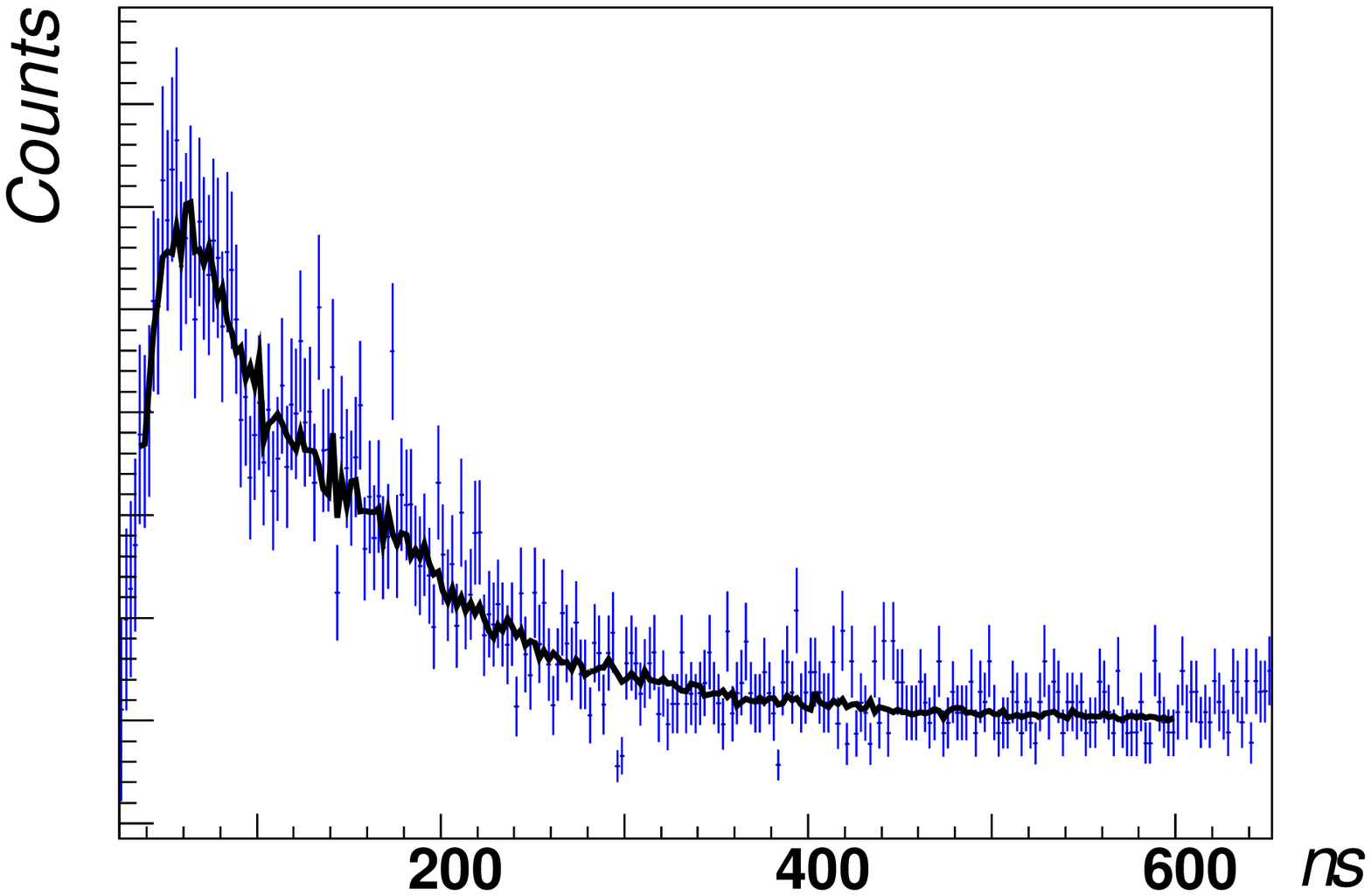}
\includegraphics[width=0.2\textwidth]{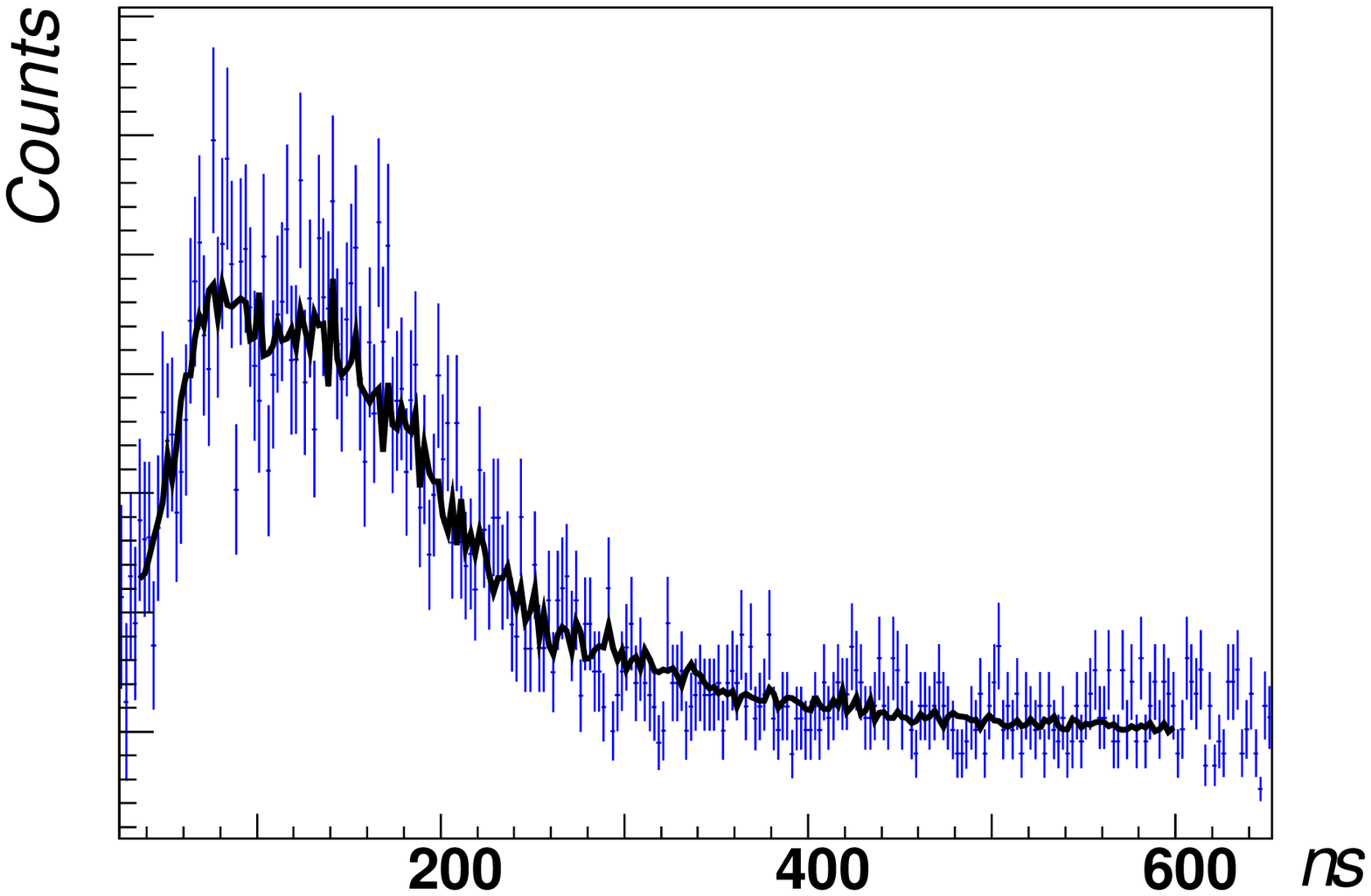}
\includegraphics[width=0.2\textwidth]{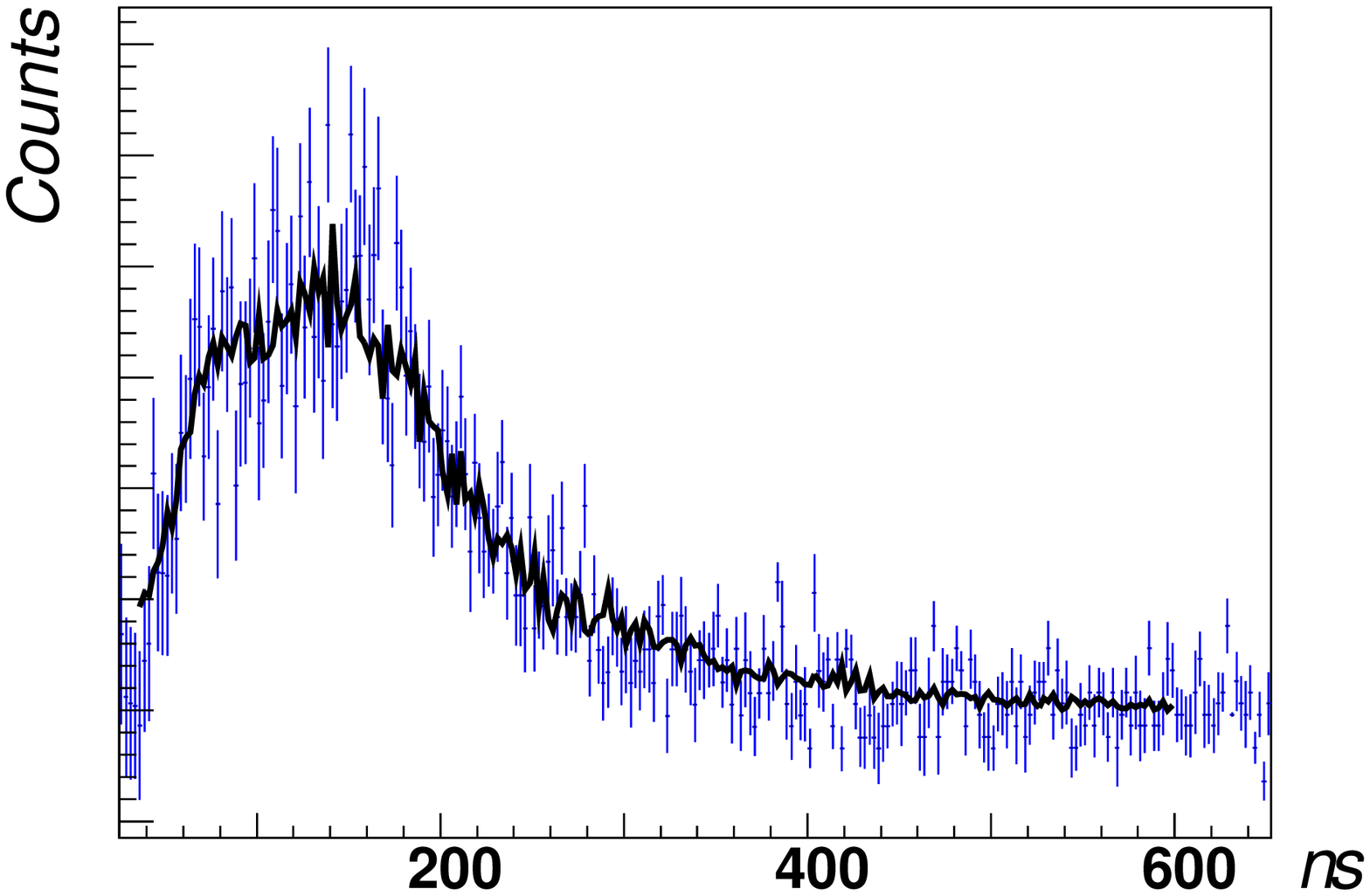}
\vspace{0.cm}
\includegraphics[width=0.2\textwidth]{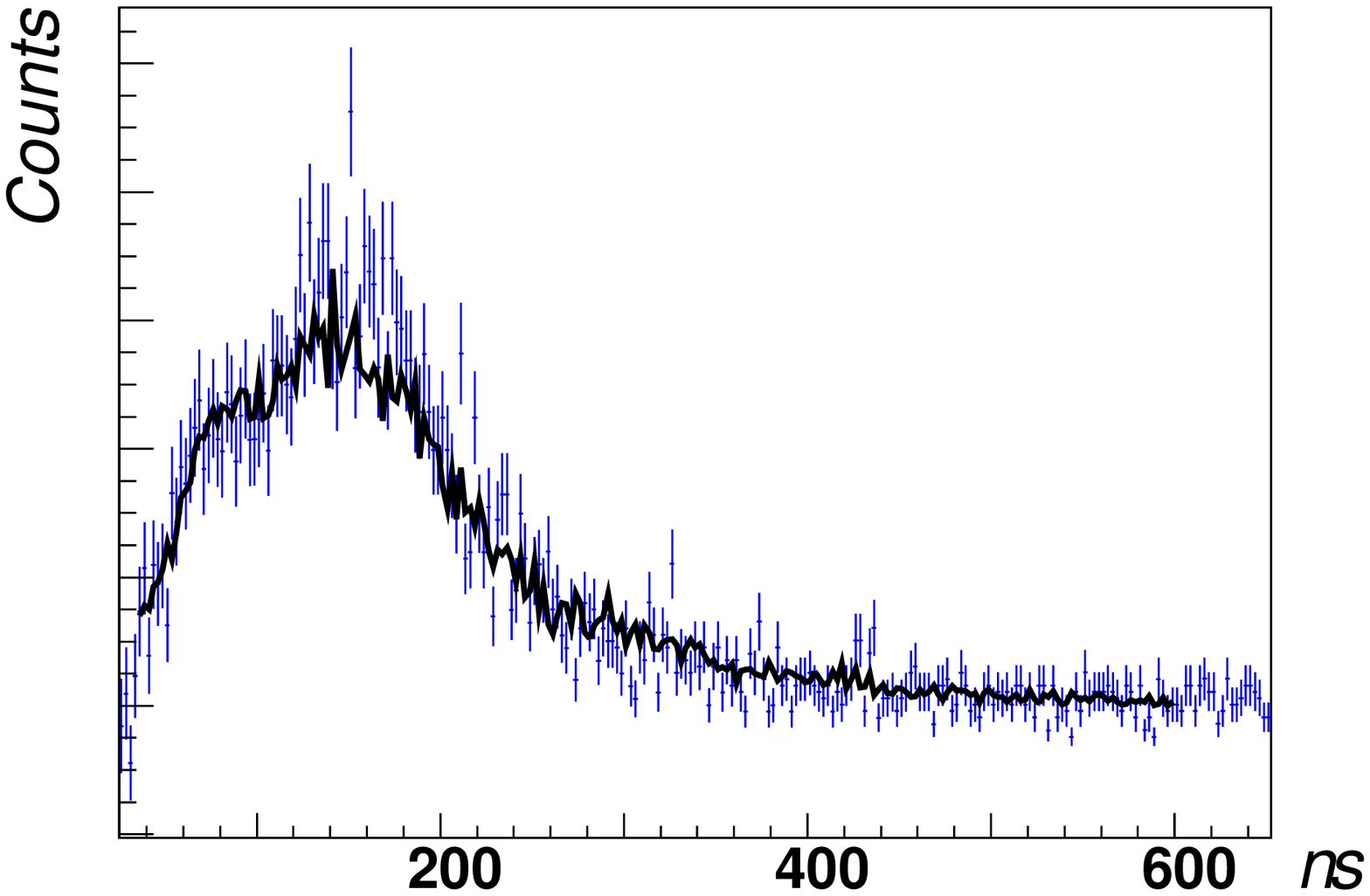}
\includegraphics[width=0.2\textwidth]{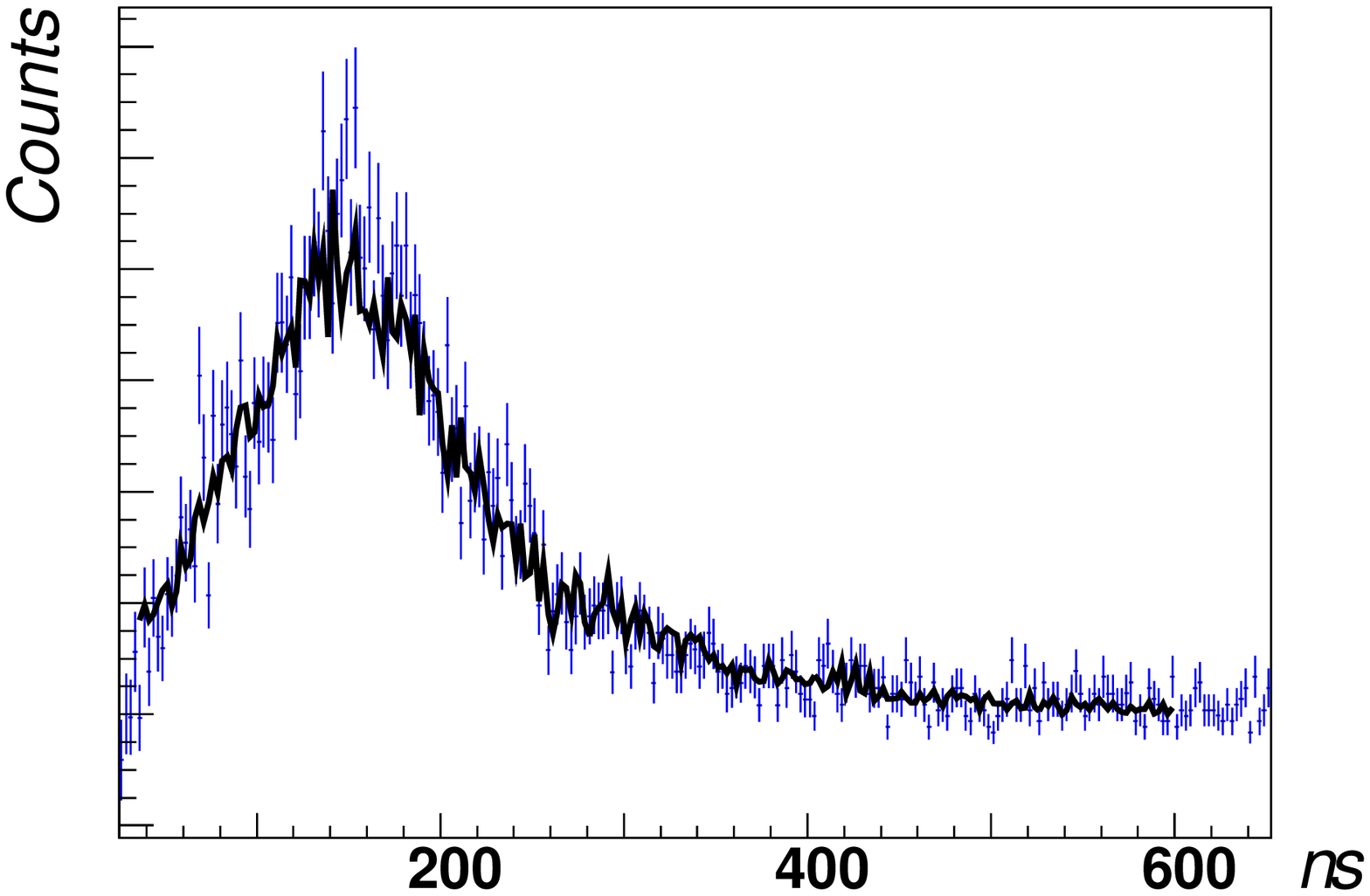}
\includegraphics[width=0.2\textwidth]{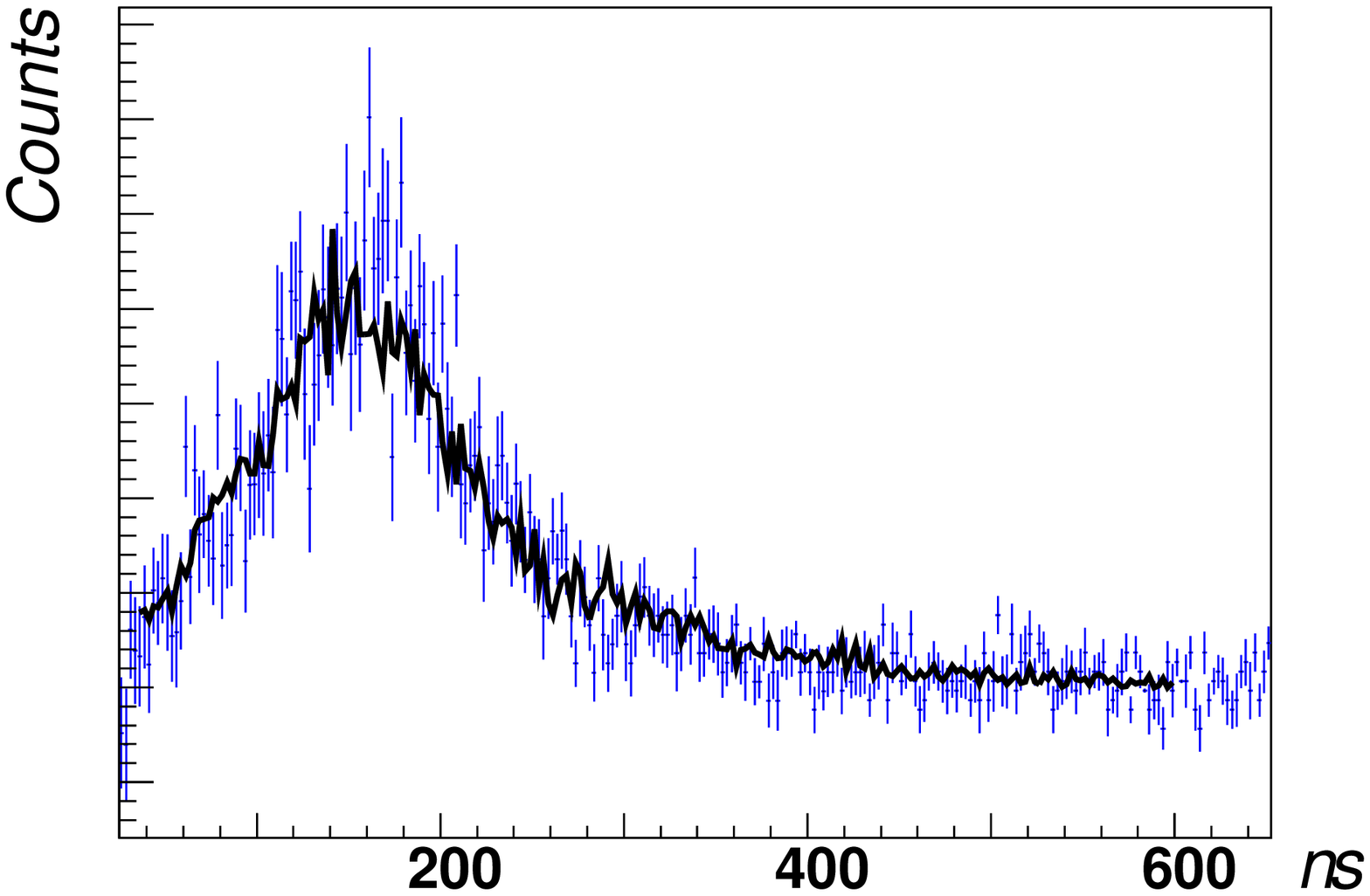}
\vspace{0.cm}
\caption {TOF spectra for MOF-5 (1,3,4,5,7.5 and 10 kV) fitted to the simulation as explained in the text. For comparison the time for a Ps emitted perpendicular to the surface with 125 meV to reach the center of the collimator slit is 150 ns.}
\label{fig:toffit}
%\end{center}
\end{figure}

\begin{figure}[h!]
%\begin{center}
\includegraphics[width=0.35\textwidth]{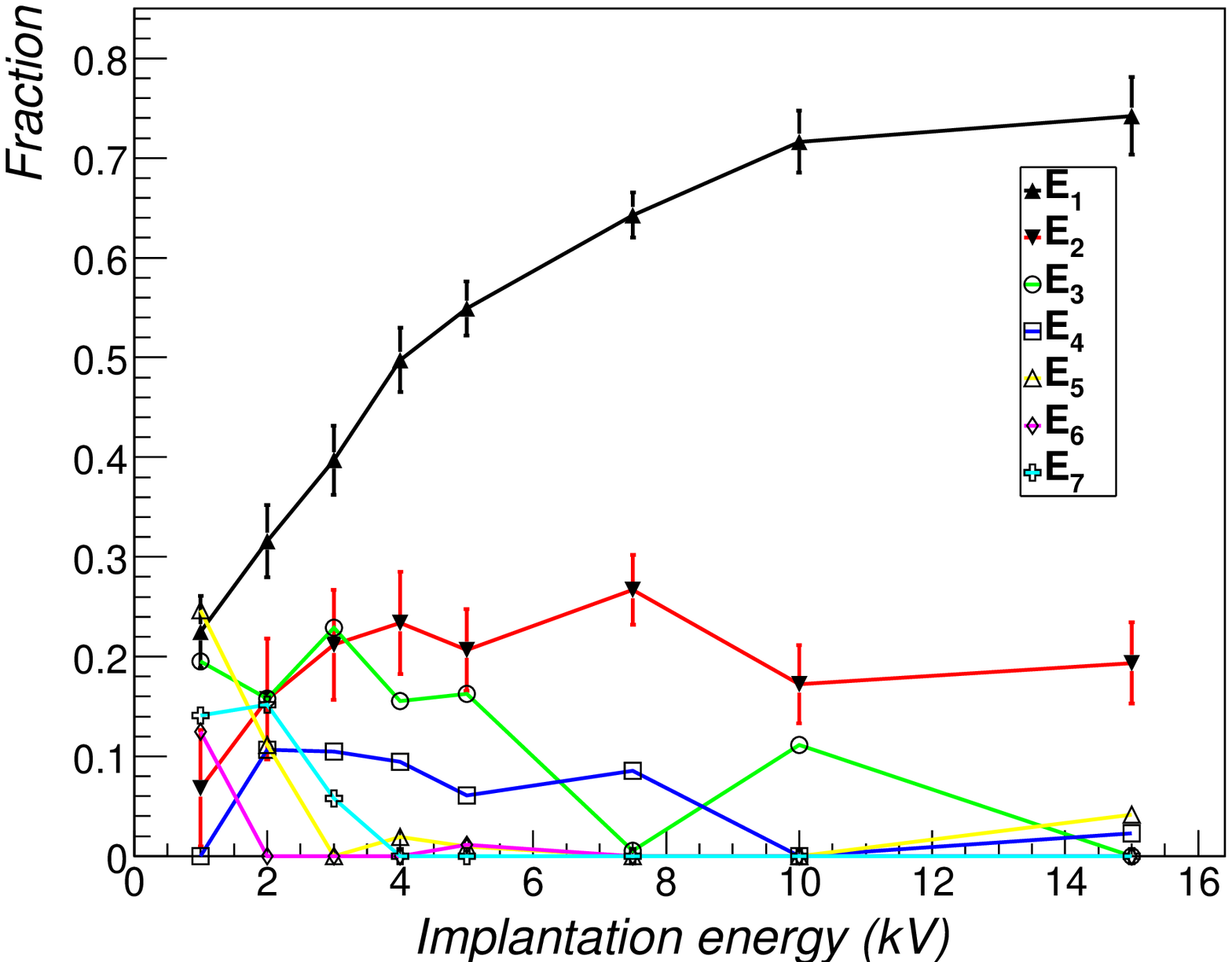}
\includegraphics[width=0.35\textwidth]{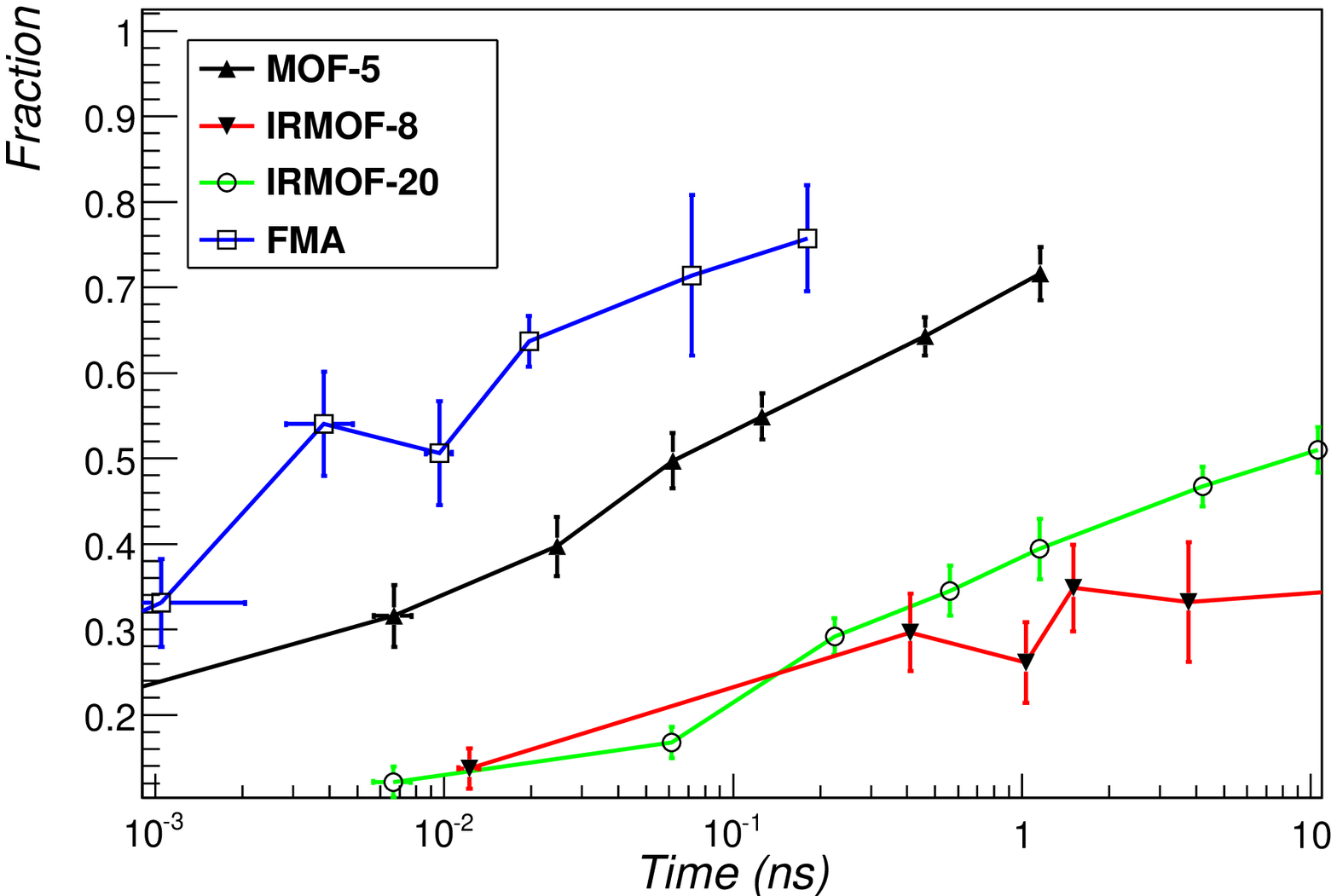}
\vspace{0.cm}
\caption{Top: evolution of the different energy components in MOF-5 as a function of the implantation energy. For readability only the error of the first two components is presented. Bottom: evolution of the Bloch state populations as a function of time for the different MOFs (see text for details).}
\label{fig:timefit}
%\end{center}
\end{figure}

To determine the emission energy of Ps into vacuum we fit the measured TOF distributions with a sum of spectra simulated with Geant4 code \cite{geant,oPsTOF}. %(see Fig. \ref{fig:tofsim}). 
We use monoenergetic distributions emitted isotropically with their fractions as the free parameters. Our choice is justified by the fact that Ps in the MOF pores can only occupy discrete energy levels. Higher emission energies (corresponding to unbound Ps) are not well resolved by the TOF detector, and so may be approximated by a sum of several mono-energetic distributions. The best fits at the lowest e$^+$ implantation energies are achieved with 6 or 7 energy distributions (see Table \ref{tbl3}). At high energies e$^+$ (7.5--10 kV) only 2 or 3 of the lowest Ps energy components are found.
The best fits give the same values within the quoted errors for all the different e$^+$ implantation energies we measured. Typical reduced $\chi^2$ are of the order of 1--1.15 for 230 d.o.f. depending on the e$^+$ implantation energy. As a cross check we took data at different slit positions (10, 15, 20 mm) achieving consistent results. From these measurements, one can also determine the delay time for Ps emission into vacuum due to the diffusion to the surface \cite{CassidyDelayed, BrusaDelayed}. This can be estimated by extrapolation to a slit position of 0 mm. We obtain values in the range from few ns up to 15 ns depending on the sample for implantation energies of 10 kV. We do not correct for this time (few \% to 10\% of the TOF time) in the determination of the emission energy but this effect is included in the quoted error.
Table \ref{tbl3} summarizes the observed energy components for the four MOFs.
In Fig. \ref{fig:toffit} we show the TOF spectra of MOF-5 for different e$^+$ implantation energies. In Fig. \ref{fig:timefit}, one can see the evolution of the different energies found in the fitting as a function of the implantation energy.
%\begin{figure}[h!]
%\centering
%\includegraphics[width=\columnwidth]{Plots/TOF_MOF5_3800VAndSim.eps}
%\vspace{0.cm}
%\caption{Simulated spectra and data.}
%\label{fig:tofsim}
%\end{figure}
% 
%For Ps localized in 3D boxes of 1-1.5 nm, the minimal energy that one would expect to measure is the ground state energy which should be of the order of $\simeq$200-400 meV. 
%However our measurements clearly show that there is a lower energy components which population increases with the time Ps spends in the MOFs (see Fig.\ref{fig:timefit}).
%\begin{figure}[h!]
%\begin{center}
%\includegraphics[width=0.4\textwidth]{Plots/MOFsBSOccupationvsTime.eps}
%\vspace{0.cm}
%\caption{Evolution of the Bloch state populations as a function of time for the different MOFs.}
%\label{fig:BSvst}
%\end{center}
%\end{figure}
The lowest energy component evident in Fig. \ref{fig:timefit} can be understood in terms of Ps in a Bloch state. This can be demonstrated calculating the energy of Ps in the first band using Eq. \ref{eq:kp}. The closest H-H distances listed in Table \ref{tbl1} can be seen as the lowest bound for $a$. We would expect the mean size of $a$ to be between this value and the cluster-to-cluster distance. To reproduce the experimental results (Table \ref{tbl3}), i.e. energy of Ps in the Bloch state one needs $a$ to be as listed in the second column of Table \ref{tbl4}.
 For all the MOFs, the values of $a$ lie within the range given above and after subtraction of twice the H van der Waals radii they are in good agreement with the H-H distances reported in Table \ref{tbl1} as we would expect for our model to be valid. 
In the other columns of Table \ref{tbl4}, those values are used to calculate with Eq. \ref{eq:Elevels} the possible bound states of Ps in a potential well of side $a$ and height $V_0$. From the comparison of Tables \ref{tbl3} and \ref{tbl4}, it looks like our measurements cannot resolve the $E_{111}$ and the $E_{211}$ levels but those appear to be merged in the $E_2$ component. However, the  $E_{221}$ and the $E_{222}$ energy levels seem to be well reproduced by the $E_3$ and $E_4$ components for all the measured samples. Above $E_4$ no bound state can exist and therefore we identify the sum of $E_5$, $E_6$ and $E_7$ components as an approximation of the continuum. Fig. \ref{fig:timefit} can thus be viewed as the evolution with time of the population in the delocalized ($E_1$), the localized states ($E_2$-$E_4$) and the continuum ($E_5$-$E_7$).
In Fig. \ref{fig:timefit}, we plot the occupation of the Bloch state as a function of the time $t$ Ps spent in the films. This is calculated using the values reported in Table \ref{tbl2} for the mean implantation depths and the lifetime of Ps in the MOFs measured with the BaF$_2$ spectrometer (see Eq. 23 of \cite{Cassidy2010}).
\begin{center}
\begin{table*}
\caption{Observed energy components (in meV) for the different MOFs.}\label{tbl3}
\small{
\hfill{}
\begin{ruledtabular}
\begin{tabular}{ccccccccc}
 Sample & $E_1$ & $E_2$ & $E_3$ & $E_4$ & $E_5$ & $E_6$ & $E_7$\\
\hline
IRMOF-20  &  120$\pm$10  & 250$\pm$50 & 500$\pm$50 & 700$\pm$50 & 900$\pm$50  & 1100$\pm$50  & 1500$\pm$100 & \\
IRMOF-8   &  138$\pm$15  & 300$\pm$50 & 600$\pm$50 & 800$\pm$50 & 1200$\pm$50  & 1400$\pm$50  & 1800$\pm$100 & \\
MOF-5     &  125$\pm$5  & 300$\pm$50 & 600$\pm$50 & 800$\pm$50 & 1050$\pm$50 & 1300$\pm$50  & 1800$\pm$100 &  \\
FMA       &  162$\pm$5 & 400$\pm$50 & 800$\pm$50 & 1200$\pm$50 & 1500$\pm$50  & 1800$\pm$100 & 2400$\pm$100 &  \\
\end{tabular} 
\end{ruledtabular}}
\hfill{}
\end{table*}
\end{center}
\begin{center}
\begin{table}
\caption{\label{tbl4} Calculated energy levels (in meV) for the different MOF structures for a given side length $a$.}
{\small
\hfill{}
\begin{ruledtabular}
\begin{tabular}{ccccccc}
 Sample 	& $a$ (nm) 		&  BS 		& $E_{111}$ &$E_{211}$ 	&$E_{221}$	&$E_{222}$\\
\hline
IRMOF-20 	& 1.17$\pm$0.02	&  122$\pm$10 & 188 		& 350 		&  512	 	& 675	\\
IRMOF-8  	& 1.15$\pm$0.03	&  139$\pm$15 & 194 		& 362 		&  530		& 699	\\
MOF-5    	& 1.08$\pm$0.01	&  122$\pm$5 & 220		& 411 		&  603 		& 794	\\
FMA     	& 0.91$\pm$0.01	&  168$\pm$5 & 311 		& 580 		&  849		& 1118	\\
\end{tabular} 
\end{ruledtabular}}
\hfill{}
\end{table}
\end{center}
%To justify the validity of our estimate for the potential barrier height $V_0$, we have calculated the hydrogen atom potential near a benzenoid linker using the Gaussian code \cite{g09}. This potential shows a jump of approximately 0.37 eV when the molecular linker distance from the H atom varies from 0.1 to 0.3 nm along a perpendicular direction with respect the benzene ring. Moreover, one can compare the binding energies for hydrogen atoms on MOF calculated as [$E$(MOF + $N$H) - $E$(MOF)]/$N$- $E$(H) for $N$ hydrogen \cite{ganz}. For example, the binding energy is 2.8 eV/H for MOF-5 + 18H (adding six H to each benzene) while for the saturation case of MOF-5 + 30H, one finds 2.5 eV/H. Interestingly, by taking the differences of these binding energies, one also obtains an energy jump of the order of 0.3 eV. Nevertheless, it is important to keep in mind that Ps does not bind like H because of its large zero point motion.\par
In conclusion, our measurements provide  definitive evidence of Ps delocalization in MOFs.  We confirm with different experimental technique the unusually long diffusion lengths reported in \cite{Dutta2013} as evidence of delocalization. More importantly, we find that the lowest component of the Ps emission energy for all four  MOFs is significantly too low for that expected from Ps localized in the MOFs' cells.  Instead, we show that this low emission energy agrees with a calculation assuming Ps delocalized in a Kronig-Penney model potential with dimension $a$ determined from the literature for each of the different-sized MOFs.
In addition, we measured the time evolution of the population for the Ps bound and delocalized states in MOFs and show how their occupation can be controlled by tuning the e$^+$ implantation energy.
%Like Ps, an exciton (i.e. electron-hole pair) placed in a periodic lattice behaves like an extended wave, while in the presence of disorder, the exciton can localize as a classical particle. In complex systems such as green plants, excitons can experience competing effects: disorder and some interactions tend to localize individual excitons at chromophores, while the energy resonant transfer tends to delocalize the excitons. This subtle balance makes photosynthetic systems excellent resources for studying and imitating efficient energy transfer. In particular, the resonant energy transfer  provides an efficient avenue for energy transfer between molecules separated by distances of up to 10 nm within an interpenetrating networks. 
The present measurements represent a clear, textbook example of a particle in a finite well and an application of the Bloch theorem to the Kronig-Penney potential, as applied to Ps in crystalline, microporous materials. In particular, our system could be used to simulate exciton migration in molecular crystals, which can be either wavelike (coherent) or diffusionlike (incoherent) \cite{grover}.
Such analogies  between excitons and Ps has been already investigated in the alkali halides by Hyodo and Bondarev \cite{bondarev}. However, since in the case of the MOF's the Ps lives much longer and diffuses over relatively large distances (10 $\mu$m), the migration study is considerably facilitated.
Moreover, from a practical point of view, the relative populations of Bloch state and localized Ps might be used to ascertain the extent of interconnectivity in a porous framework. This may be especially useful to quantify occluded guests or local pore collapse, inhomogeneities notoriously difficult to detect by common techniques. Extending this work to the study of MOFs with different pore geometries (i.e. 1D channel pores) may lead to a deeper understanding of the intimate relationship between pore geometry/interconnectivity and Ps Bloch state characteristics.
 %These present measurements are not only a nice text book example of a particle in a finite well and an application of the Bloch theorem to the Kronig-Penney potential but they could hopefully help to get a better understanding of diffusion of molecules in MOFs. 
%shedding some light in the debated question on the apertures. 
%Ps shows great promise as a unique probe of these new crystals. 
%Furthermore This is a first step towards a Ps condensate.

This work was supported by the Swiss National Science Foundation (grant PZ00P2\_132059) and ETH Zurich (grant ETH-47-12-1). B. Barbiellini is supported by the DOE grants No. DE-FG02-07ER46352 and No. DE-AC02-05CH11231 for theory support at the Advanced Light Source, Berkeley and the allocation of supercomputer time at NERSC. MOF's synthesis was supported by the DOE (DE-SC0004888). We are very grateful to A. Rubbia, K. Kirch, G. Dissertori and the IPP at ETH Zurich for their essential support, to S. Eijt for very useful discussions, to A. Gendotti and the IPP workshop for their help with the construction.


\begin{thebibliography}{99}

\bibitem{Deutsch1} M. Deutsch, Phys. Rev. 82, 455 (1951)  

\bibitem{SavelyPhysRep}
S.~G.~Karshenboim,
  %``Precision physics of simple atoms: QED tests, nuclear structure and
  %fundamental constants,''
  Phys.\ Rept.\ 422, 1 (2005).
 %
\bibitem{oPsNewPhysics}
 A. Badertscher, P. Crivelli, W. Fetscher, U. Gendotti, S. Gninenko, V. Postoev, A. Rubbia, V. Samoylenko, D. Sillou, Phys. Rev. D 75, 032004 (2007).
\bibitem{Mills1981}
A. P. Mills, Phys. Rev. Lett. 46, 717 (1981).
\bibitem{Cassidy2007}
D. B. Cassidy and A. P. Mills, Nature 449, 195 (2007).
\bibitem{barbiellini} B. Barbiellini and P.M. Platzman, Phys. Stat. Sol. C6, 2523 (2009).
\bibitem{gidley1} D. W. Gidley, H. G. Peng, R. S. Vallery Annual Review of Materials Research 36, 49 (2006).
\bibitem{brandt} W. Brandt, G. Coussot, and R. Paulin, Phys. Rev. Lett. 23, 522 (1969).
\bibitem{Greenberger} A Greenberger, A.P. Mills, A. Thompson, and S.  Berko, Phys. Lett. 32A, 72 (1970).
\bibitem{inoue}
K. Inoue, N. Suzuki, I. V. Bondarev, and T. Hyodo, Phys.Rev. B 76, 024304 (2007), and the references therein.
\bibitem{nagai}
 Nagai et al. Phys. Rev. B 60, 7677 (1999).
\bibitem{Dutta2013}
D. Dutta, J. I. Feldblyum, D. W. Gidley, J. Imirzian, M. Liu, A. J. Matzger, R. S. Vallery and A. G. Wong-Foy, Phys. Rev. Lett. 110, 197403 (2013).
\bibitem{janiak}
C. Janiak, J. K. Vietha, New J. Chem. 34, 2366 (2010).
\bibitem{farha}
O. K. Farha, I. Eryazici, N. C. Jeong, B. G. Hauser, C. E. Wilmer, A. A. Sarjeant, R. Q. Snurr, S. T. Nguyen, A. O. Yazaydın, and J. T. Hupp, J. Am. Chem. Soc.,134 (36), 15016–15021 (2012).
\bibitem{Makal2012}
T. A. Makal, J.-R. Li, W. Lu and H.-C. Zhou, Chem. Soc. Rev. 41, 7761 (2012).
\bibitem{fma}
M. Xue, Y. Liu, R.M. Schaffino, S. Xiang, X. Zhao, G. Zhu, S. Qiu, B. Chen,
Inorg. Chem., 48 (11), 4649 (2009).
\bibitem{mof5}
H. Li, M. Eddaoudi, M. O'Keeffe, O. M. Yaghi, Nature 402, 276 (1999).
\bibitem{irmof8}
M. Eddaoudi, J. Kim, N. Rosi, D. Vodak, J. Wachter, M. O'Keeffe, O.M. Yaghi, Science 295, 469 (2002). 
\bibitem{irmof20}
J. L. C. Rowsell and O. M. Yaghi, J. Am. Chem. Soc., 128 (4), 1304–1315 (2006).
\bibitem{Alberola2006}
N. Alberola, T. Anthonioz, A. Badertscher, C. Bas, A. S. Belov, P. Crivelli, S. N. Gninenko, N. A. Golubev, M. M. Kirsanov, A. Rubbia and D. Sillou, Nucl. Instr. Meth. Phys. Res. A 560, 224 (2006).
\bibitem{oPsTOF}
 P. Crivelli, U. Gendotti, A. Rubbia, L. Liszkay, P. Perez, C. Corbel, Phys. Rev. A 81, 052703 (2010).

\bibitem{gidley2} D. W. Gidley, W. E. Frieze, T. L. Dull, A. F. Yee, E. T. Ryan, and H. M. Ho, Phys. Rev. B 60, R5157 (1999).
\bibitem{gidley3} D. W. Gidley, W. E. Frieze, T. L. Dull, J. Sun, A. F. Yee, C. V. Nguyen, and D. Y. Yoon, Appl. Phys. Lett. 76, 1282 (2000).

\bibitem{Makhovian}
A. P. Mills Jr., R. J. Wilson, Phys. Rev. A 26, 490 (1982).
\bibitem{Batista}
M. Batista, M. Lakner, J. Peternelj, Eur. J. Phys. 25, 145 (2004).
\bibitem{MuoniumPRL}
A. Antognini, P. Crivelli, T. Prokscha,‡, K. S. Khaw, B. Barbiellini, L. Liszkay, K. Kirch, K. Kwuida, E. Morenzoni, F. M. Piegsa, Z. Salman, and A. Suter, Phys. Rev. Lett. 108, 143401 (2012). 
\bibitem{Cassidy2010}
D.B. Cassidy, P. Crivelli, T.H. Hisakado, L. Liszkay, V.E. Meligne, P. Perez, H.W.K. Tom, A.P. Mills, Phys. Rev. A 81, 012715 (2010).
 \bibitem{bondarev}
I.V. Bondarev and T. Hyodo, PRB 57, 11341 (1998).
 \bibitem{Anderson1958}
P. W. Anderson, Phys. Rev. 109, 1492 (1958).
 \bibitem{baugher}
A. H. Baugher, W. J. Kossler, K. G. Petzinger, Macromolecules 29, 7280 (1996).
\bibitem{OurAPL} L. Liszkay et al., Appl. Phys. Lett. 95, 124103 (2009).
\bibitem{geant} Agostinelli et al., Nucl. Instr. Method A 506, 250 (2003).
\bibitem{CassidyDelayed}
D.B. Cassidy, T.H. Hisakado, V.E. Meligne, H.W.K. Tom, A.P. Mills, Jr., Phys. Rev. A 82, 052511 (2010).
\bibitem{BrusaDelayed}
L. Di Noto, S. Mariazzi, M. Bettonte, G. Nebbia, R.S. Brusa, Eur. Phys. J. D 66, 118 (2012).  
%\bibitem{g09}
%M. J. Frisch {\em et al.} Gaussian~09, Revision A.1, Gaussian Inc., Wallingford, CT (2009).
%\bibitem{ganz}
%E. Ganz and M. Dornfeld, J. Phys. Chem. 116, 3661 (2012).
\bibitem{grover}
M. Grover and R. Silbey, J. Phys. Chem. 54, 4843 (1971).


\end{thebibliography}
\end{document}